\documentclass[twocolumn]{aastex63}

\newcommand\rp{\emph{r}-process}

\usepackage{amssymb}
\usepackage{enumerate}
\usepackage{graphicx}
\usepackage{amsmath}
\usepackage{float}
\usepackage{xcolor}
\usepackage{hyperref}
\hypersetup{
    bookmarks=true,                 
    unicode=false,                  
    pdftoolbar=true,                
    pdfmenubar=true,                
    pdffitwindow=true,              
    pdfstartview={FitH},            
    pdftitle={Yields of Milky Way Mergers}, 
    pdfauthor={eholmbeck},          
    pdfsubject={Astrophysics},    
    pdfcreator={dvipdf},            
    pdfproducer={dvipdf},           
    pdfkeywords={r-process, stars, neutron stars, equation of state, EOS, elements}, 
    pdfnewwindow=true,              
    colorlinks=true,                
    linkcolor=magenta,              
    citecolor=teal,                 
    filecolor=magenta,              
    urlcolor=cyan,                  
    breaklinks=true,
    linktocpage
}

\newcommand\Msun{M$_\sun$}

\graphicspath{{Plots/}{./}{arXiv}}

\received{\today}
\revised{}
\accepted{}

\shorttitle{Yields of Milky Way Mergers}
\shortauthors{Holmbeck \& Andrews}

\begin{document}

\title{Total \emph{r}-process Yields of Milky Way Neutron Star Mergers}

\correspondingauthor{Erika M.\ Holmbeck}
\email{eholmbeck@carnegiescience.edu}

\author[0000-0002-5463-6800]{Erika M.\ Holmbeck}
\altaffiliation{NHFP Hubble Fellow}
\affiliation{Observatories of the Carnegie Institution for Science, 813 Santa Barbara Street, Pasadena, CA 91101, USA}

\author[0000-0001-5261-3923]{Jeff J.\ Andrews}
\affiliation{Department of Physics, University of Florida, 2001 Museum Road, Gainesville, FL 32611, USA}
\affiliation{Institute for Fundamental Theory, University of Florida, 2001 Museum Road, Gainesville, FL 32611, USA}
\affiliation{Center for Interdisciplinary Exploration and Research in Astrophysics (CIERA), 1800 Sherman Avenue, Evanston, IL 60201, USA}

\begin{abstract}
While it is now known that double neutron star binary systems (DNSs) are copious producers of heavy elements, there remains much speculation about whether they are the sole or even principal site of rapid neutron-capture (\rp) nucleosynthesis, one of the primary ways in which heavy elements are produced.
The occurrence rates, delay times, and galactic environments of DNSs hold sway over estimating their total contribution to the elemental abundances in the Solar system and the Galaxy.
Furthermore, the expected elemental yield for DNSs may depend on the merger parameters themselves---such as their stellar masses and radii---which is not currently considered in many galactic chemical evolution models.
Using the characteristics of the observed sample of DNSs in the Milky Way as a guide, we predict the expected nucleosynthetic yields that a population of DNSs would produce upon merger, and we compare that nucleosynthetic signature to the heavy-element abundance pattern of the Solar system elements.
We find that with our current models, the present DNS population favors production of the lighter \rp\ elements, while underproducing the heaviest elements relative to the Solar system.
This inconsistency could imply an additional site for the heaviest elements or a population of DNSs much different from that observed today.
\end{abstract}

\keywords{Nucleosynthesis (1131), R-process (1324), Neutron stars (1108), Chemical abundances (224), Galaxy chemical evolution (580)}


\section{Introduction}
\label{sec:intro}

The trans-iron elements in the Solar system attribute the vast majority of their origins to neutron capture.
Neutron capture is further divided into two main sub-classes, slow (``\emph{s}") and rapid (``\emph{r}"), generally reflecting the neutron density of the environment in which each takes place \citep{Burbidge1957,Cameron1957}.
Requiring extreme neutron densities, the astrophysical site of the \rp\ has been a topic of debate for decades, with compact object mergers being proposed \rp\ sites since \citet{lattimer1974}.
Since these early works, many studies have continued to claim that compact object mergers---particularly DNSs---can be responsible for the \rp\ \citep{lattimer1976,symbalisty1982,eichler1989,Freiburghaus1999}.
However, it was only recently that this theory has been observationally confirmed through the electromagnetic signature following the gravitational-wave detection of a binary neutron-star (NS) merger (NSM) \citep[GW170817;][]{GW170817_detection,GW170817_followup}.
Although the electromagnetic light curve and spectrum (AT2017gfo) associated with this event suggests that the \rp\ took place \citep{coulter2017,Cowperthwaite2017,watson2019, metzger2019, margutti2021}, it is still unknown whether NSMs can account for the majority of the abundance of \rp\ elements observed in the Solar system and throughout the Galaxy \citep[see][for a review]{siegel2022}.
Difficulties with the NSM model include explaining the \rp\ enrichment of ultra-faint dwarf galaxies with escape velocities far below the typical natal kick velocity applied to newborn NSs \citep{beniamini2016, safarzadeh2019}, the [Eu/Fe] abundances of metal-rich stars in the Milky Way disk \citep{cote2019}, and inconsistencies between the observed lanthanide fraction of GW170817 and that of \rp\ enhanced stars in the Milky Way halo \citep{ji2019}.
Proposed additional \rp\ production sites include magnetorotational core-collapse supernova \citep{winteler2012,nishimura2015}, magnetized winds from proto-neutron stars \citep{thompson2003}, supernova fallback \citep{fryer2006}, and accretion disk winds from collapsars \citep{surman2008,siegel2018}.

The study of NSMs in a Galactic context is a two-pronged topic.
First is the hydrodynamical investigation of the NSM itself and the detailed generation of \rp\ elements \citep{Freiburghaus1999, metzger2010, roberts2011, kasen2013, rosswog2014, grossman2014}.
Accurate simulations require a detailed treatment of hydrodynamics including general relativistic and radiative effects as well as a complex nuclear reaction network \citep[for a review, see][]{metzger2019}. 
Second is the mixing processes in the interstellar medium and the broader chemical evolution of the Galaxy \citep{argast2004, ishimaru2015, van_de_voort2015, cote2017, macias2018, ojima2018, macias2019, beniamini2020, van_de_voort2020, wanajo2021}.
Studies disagree on the ability of turbulent mixing to efficiently bring \rp\ enriched material back to the Milky Way disk \citep[e.g.,][]{amend2022}.
However, these studies clearly show that the \rp\ site in the disk must be much rarer than normal core-collapse SN, consistent with NSM \citep{macias2018, macias2019}, and that metal-rich stars such as the Sun must have been enriched by a large number of individual \rp\ events, so that stochastic effects can be ignored.

In this work, we assume the \rp\ abundance pattern of the Sun is representative of that of the Milky Way disk as a whole. 
Previously, \citet{Freiburghaus1999} and \citet{metzger2010} have compared spectroscopic abundances of a variety of DNS merger simulations with the Solar abundance, finding reasonable agreement.
However, the nucleosynthetic yield of NSMs depends on a combination of factors including the two components' masses, the NS Equation of State (EOS), and the various models for the contribution from disk winds and dynamical ejecta.
While the latter three may be uncertain, the known population of DNSs in the Milky Way inform us about the masses of merging DNSs. 
Here, we combine a variety of merger simulations with a realistic population of DNSs informed by the observed Milky Way systems, accounting for those systems that have already merged and are not seen today, to determine a realistic nucleosynthetic yield of \rp\ elements at the present day.
This study differs from previous studies of Galactic chemical evolution in that we explore NSM yields that depend on the NS masses and radii.
We also systematically explore the effect of nuclear uncertainties on our NSM yields.
We begin our study by calculating the \rp\ abundances that could be expected from the Milky Way population of DNSs, starting with an overview of the observed sample of DNSs and its nuances in Section~\ref{sec:DNS}.
We then describe our procedure for calculating nucleosynthetic yields from a DNS merger with a particular combination of NS masses in Section~\ref{sec:method}.
In Section~\ref{sec:results} we combine the DNS sample with our model of elemental yields to derive a nucleosynthetic fingerprint from a Galaxy-scale population of NSMs.
Finally, we provide some conclusions in Section~\ref{sec:conclusions}.


\begin{deluxetable*}{l c c c c c c c}
\tablecaption{Observed DNS systems and their ejecta properties for our fiducial model. The median and 16th and 84th percentiles between all random samples is shown for the wind and dynamical ejecta masses ($m_i$) and their respective mass fractions of lanthanides and actinides ($X^{\rm lan}_i$).\label{tab:results}}
\tablehead{
\colhead{Name} & \colhead{$M_1$} & \colhead{$M_2$} & \colhead{Weight} & \colhead{$m_{\rm wind}$} & \colhead{$m_{\rm dyn}$} & \colhead{$X^{\rm lan}_{\rm wind}$} & \colhead{$X^{\rm lan}_{\rm dyn}$}
\\
\colhead{} & \colhead{(M$_\odot$)} & \colhead{(M$_\odot$)} & \colhead{} & \colhead{(10$^{-3}$\,M$_\odot$)} & \colhead{(10$^{-3}$\,M$_\odot$)} & \colhead{($\times$10$^{-3}$)} & \colhead{($\times$10$^{-3}$)}
}
\startdata
\multicolumn{8}{c}{Field Systems}\\
\hline
J0737$-$3039	& $1.3381\pm 0.0007$ & $1.2489\pm 0.0007$ & 4.8 & $176.8_{-83.0}^{+85.9}$ & $2.7_{-1.3}^{+1.3}$ & $1.9_{-0.0}^{+0.0}$ & $148.7_{-11.0}^{+5.2}$  \\
B1534+12	    & $1.3452\pm 0.0010$ & $1.3332\pm 0.0010$ & 0.6 & $164.5_{-78.2}^{+81.0}$ & $3.5_{-1.7}^{+1.8}$ & $1.9_{-0.0}^{+0.0}$ & $34.5_{-22.0}^{+33.7}$\\
J1756$-$2251	& $1.312\pm 0.017$	& $1.258\pm 0.017$ & 0.6 & $171.8_{-80.4}^{+83.8}$ & $2.1_{-1.1}^{+1.2}$ & $1.9_{-0.0}^{+0.0}$ & $140.7_{-18.5}^{+12.3}$ \\
J1906+0746	    & $1.322\pm 0.011$	& $1.291\pm 0.011$ & 2.2 & $168.7_{-79.4}^{+82.5}$ & $2.6_{-1.3}^{+1.3}$ & $1.9_{-0.0}^{+0.0}$ & $110.8_{-31.9}^{+35.6}$ \\
B1913+16	    &  $1.4398\pm 0.0002$	& $1.3886\pm 0.0002$ & 6.5 & $157.0_{-88.1}^{+85.4}$ & $5.2_{-2.4}^{+2.5}$ & $1.9_{-0.0}^{+0.0}$ & $15.8_{-10.2}^{+25.5}$ \\
J1913+1102	    & $1.62\pm 0.03$	& $1.27\pm 0.03$ & 0.5 & $227.7_{-151.1}^{+143.4}$ & $11.4_{-5.4}^{+5.9}$ & $1.9_{-0.0}^{+0.0}$ & $137.4_{-38.0}^{+15.5}$
 \\
J1757$-$1854	& $1.3946\pm 0.0009$	& $1.3384\pm 0.0009$ & 7.8 & $168.5_{-82.4}^{+83.3}$ & $4.3_{-2.1}^{+2.2}$ & $1.9_{-0.0}^{+0.0}$ & $25.5_{-16.7}^{+32.2}$ \\
\hline
\multicolumn{8}{c}{Globular Cluster System}\\
\hline
B2127+11	& $1.358\pm 0.010$	& $1.354\pm 0.010$ & --- & $165.2_{-79.9}^{+81.0}$ & $3.9_{-1.9}^{+1.9}$ & $1.9_{-0.0}^{+0.0}$ & $20.0_{-13.2}^{+30.5}$ \\
\hline
\multicolumn{8}{c}{Gravitational Wave Detections}\\
\hline
GW170817    & $1.56\pm0.11$ & $1.22\pm 0.09$ & --- & $223.7_{-112.9}^{+140.5}$ &  $7.9_{-4.2}^{+7.3}$ & $1.9_{-0.0}^{+0.0}$ & $108.4_{-72.1}^{+37.5}$ \\
GW190425	& $2.10\pm 0.31$ & $1.43\pm0.17$ & --- & $0.2_{-0.1}^{+0.5}$  & $4.3_{-2.2}^{+3.4}$ & $44.0_{-6.1}^{+19.9}$ & $137.3_{-10.5}^{+12.0}$
\enddata
\end{deluxetable*}


\section{DNS systems}
\label{sec:DNS}

In the decades since the detection of the first DNS, the Hulse-Taylor binary \citep{hulse1975}, the number of observed systems has steadily increased.
Today, there are $\simeq$20 known or suspected DNSs in the Milky Way \citep[for a review, see][]{tauris2017}.
Of these, roughly half are expected to merge due to gravitational wave-driven orbital decay within a Hubble time.
Throughout this work, we ignore DNSs with orbits too wide to merge within a Hubble time since there is evidence that these systems formed through a different evolutionary subchannel \citep{andrews2019}.
After further restricting for only systems with well-measured masses, we arrive at a sample of seven DNS\footnote{We separately treat an eighth system, B2127$+$11, the DNS associated with the globular cluster M15, as it is expected to have been formed dynamically \citep{phinney1991a}.} which we list in Table~\ref{tab:results} along with their masses. 

Not all of the seven DNSs in our sample are equally weighted, which can be explained through two broadly defined effects \citep{narayan1991, phinney1991b, kalogera2001, kalogera2004}.
First, the shorter-period systems merge quickly, such that for every system observed, many more have already formed and merged.
A complete analysis accounts for this effect by weighting each system by the inverse of the DNS lifetime, estimated to be the sum of the pulsar characteristic age and the time to merger. (This weighting scheme also makes apparent why systems that take more than a Hubble time to merge have a negligible contribution to the characteristics of the merging DNS population.)
Second, observational biases need to be taken into account.
For example, intrinsically fainter systems with a smaller detectable volume contribute more significantly, as each detected system implies a larger underlying population.
Properly incorporating observability requires a detailed treatment of pulsar lifetimes, beaming fractions, pulsar luminosities, and the footprints and depths of pulsar surveys within which they were discovered.
A quantitative analysis combining all of these effects for the seven DNSs in our sample was undertaken by \citet{pol2019, pol2020}.
Although additional systems have been detected since these works \citep[e.g.,][]{sengar2022}, none will merge within a Hubble time.

In Table~\ref{tab:results} we include the relative weighting of each DNS which is based on the derived merger rate reported by \citet{pol2019}.
Four of the seven systems (J0737$-$3039, J1906$+$0746, B1913$+$16, and J1757$-$1854) have a significantly higher weight, due principally to the larger differences in gravitational wave-driven merger times. 

We additionally consider B2127$+$11 (also known as M15C), which is found within the core-collapsed globular cluster M15 and therefore likely to be formed dynamically.
Although this system will merge in a Hubble time, dynamically formed systems are not expected to contribute significantly to the overall merger rate of DNSs in the universe; decades of focused radio observations of globular clusters implies limited numbers of such systems, in agreement with theoretical models \citep{ye2019, ye2020}.
We treat this system separately in our subsequent analysis, removing it from any extrapolation to the \rp\ abundance pattern of the Milky Way.

Finally, we include the two DNSs detected by the LIGO-Virgo network of gravitational wave observatories, GW170817 \citep{GW170817_detection} and GW190425 \citep{GW190425}.
Analysis of the gravitational waveforms leads to mass measurement accuracies of $\sim$0.1~\Msun.
In our subsequent analysis, the two stars' masses (the mass ratio in particular) dictate the resultant abundance yield.
While GW170817 has component masses consistent with the Milky Way field population, GW190425 is unique.
Its masses are sufficiently discrepant from any other known DNS leading to multiple theories about its formation \citep{romero-shaw2020, sarafzadeh2020, vigna_gomez2021}.
We consider these gravitational wave detections as well, but leave them out of any derivation of the predicted Milky Way-averaged \rp\ abundance pattern.


\section{NSM Yields}
\label{sec:method}

We consider two main sources of NSM ejecta: dynamical and disk-wind.
The total elemental yield from a merger will be a product of the two ejecta masses and their respective compositions.
We employ a forward method in which both the ejecta masses and compositions are dependent on the NS masses and the EOS.
We combine the contributions from both the dynamical ejecta and the disk wind into a total yield for each of the 10 observed DNS systems listed in Table~\ref{tab:results}.
First we describe our fiducial model in Section \ref{sec:fiducial}, then we describe how we approximate uncertainties on these models in Section \ref{sec:error_propagation}. We discuss the effects of alternatives to our fiducial prescriptions in Sections~\ref{sec:results_macro} and \ref{sec:results_micro}.
For a schematic overview of our model, see Figure~3 in \citet{Holmbeck2022}.

\subsection{Fiducial Model}
\label{sec:fiducial}

For our fiducial model, we adopt the DD2 EOS \citep{Hempel2010}.
To describe the dynamical ejecta mass, we use Equation~6 of \citet{krueger2020}, which is a fit to a series of numerical relativity simulations.
We opt for the prescription provided by \citet{dietrich2020} in Equation~S4 for the mass of the accretion disk as a function of NS masses and the EOS, which is based on a fit to a series of 73 separate numerical relativity simulations performed by different groups.
For details on why we choose these prescriptions as our default choices, see \citet{Holmbeck2022}.
We will explore the uncertainty associated with these choices in Section~\ref{sec:results_macro}.

Next, the masses of each of these components are multiplied by their respective calculated elemental composition. We use the nuclear reaction network code Portable Routines for Integrated nucleoSynthesis Modeling \citep[PRISM;][]{Sprouse2020} with the Finite Range Droplet Model \citep[FRDM2012][]{Moller2012} nuclear mass model to calculate the nucleosynthetic yield of the NSM ejecta.

For the dynamical ejecta we use a single trajectory from NSMs from the 1.4--1.4\,\Msun\ NSM simulations of S.\ Rosswog \citep{piran2013,rosswog2013} as in \citet{korobkin2012} with varying initial electron fraction, $Y_e$ \citep[see the calculations in][]{holmbeck2019,holmbeck2021}.
Rather than use a single $Y_e$ to describe the composition of the dynamical ejecta, we allow the mass to be distributed in a narrow Gaussian around a central $Y_e$ value.
This initial central $Y_e$ is chosen from the average $Y_e$ estimate from Equation~10 in \citet{nedora2021}, which is written as a function of NS mass ratio and tidal deformability.

For the disk wind component, the elemental yield is affected by the lifetime of the NSM remnant, both in the composition and the total amount of mass that is ejected. 
We take the remnant lifetime to depend on the merger parameters as in \citet{lucca2020} and represent the fraction of the disk mass that will be ejected by fitting data in \citet{metzger2014} as follows:
	\begin{equation}
    f =
    \begin{cases}
    0.20\, e^{\tau/300\,\rm{ms}} - 0.17 & \tau\leq 530\,\rm{ms} \\
    1 & \tau\geq 530\,\rm{ms}
    \end{cases},
	\label{eqn:tau}
	\end{equation}
where $\tau$ is the NSM remnant lifetime.
Then, we use the lifetime-based nucleosynthesis yields and fitting formulae as in \citet{holmbeck2021} to obtain a lifetime-dependent composition.
In this way, a longer-lived merger remnant will eject more mass and be generally less neutron-rich than its short-lived counterpart.

Once both the component ejecta masses ($M_{\rm ej}$) and yields ($Y$) are calculated, they are simply multiplied together to obtain a total abundance yield.

\subsection{Model Uncertainties and Error Propagation}
\label{sec:error_propagation}

Significant uncertainty surrounds each step in our forward model.
We have chosen reasonable prescriptions for the dynamical and wind ejecta masses and compositions, but they are certainly not the only plausible descriptions for what are intrinsically complex, multi-scale physical processes.
We therefore estimate the uncertainty in our model taking a Monte Carlo approach; when calculating the expected yields for each NSM, we take 50,000 random samples overall to account for the effects of multiple separate sources of uncertainty.

First, for the Galactic and globular cluster DNS pairs, we take 50 random draws for the two NS masses from a normal distribution with centroids and standard deviations given by the values in Table~\ref{tab:results}.
Because the NS masses in the Galactic DNS systems are measured with high precision, the systematic uncertainty on the elemental yields from these measurements is a negligible effect.
However, the NS masses for the for the gravitational wave-detected systems are a significant source of uncertainty.
For the two systems GW170817 and GW190425, instead of sampling from a normal distribution, we take 50 samples from their posterior mass distributions: the low-spin prior mass pairs from \citet{GWTC-1}\footnote{\url{https://dcc.ligo.org/LIGO-P1800370/public}} and \citet{GW190425}\footnote{\url{https://dcc.ligo.org/LIGO-P2000026/public}}, respectively.


 	\begin{figure*}[t]
 	\centering
	\includegraphics[width=\columnwidth]{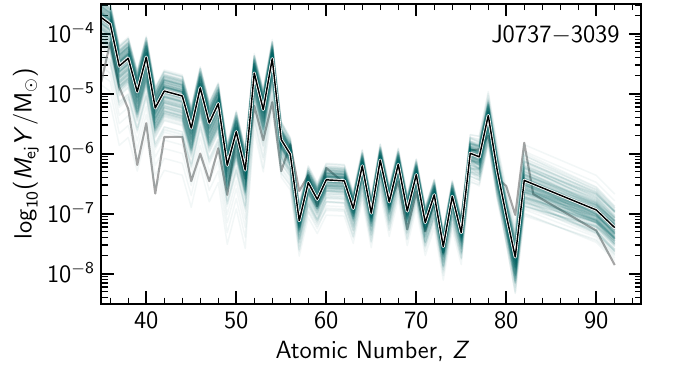}
	\includegraphics[width=\columnwidth]{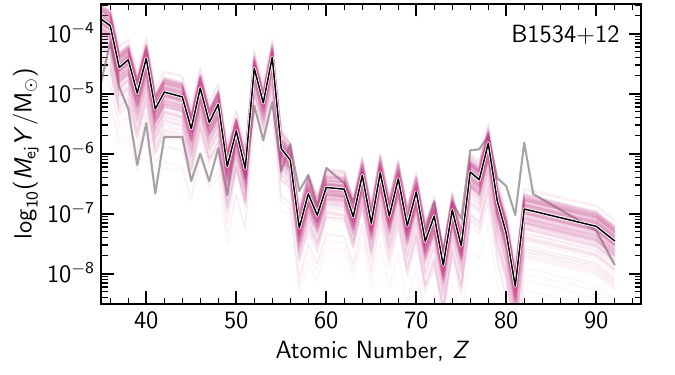}\\
	\includegraphics[width=\columnwidth]{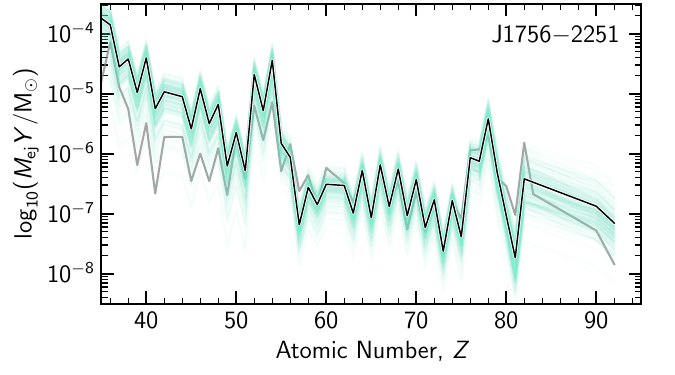}
	\includegraphics[width=\columnwidth]{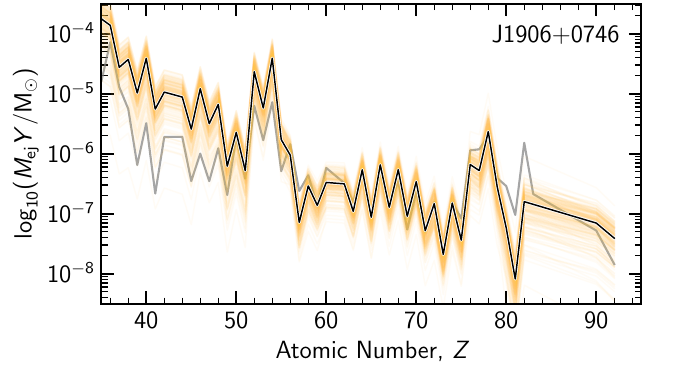}\\
	\includegraphics[width=\columnwidth]{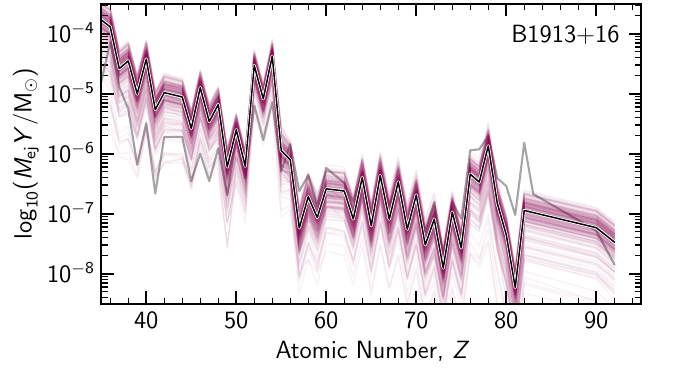}
	\includegraphics[width=\columnwidth]{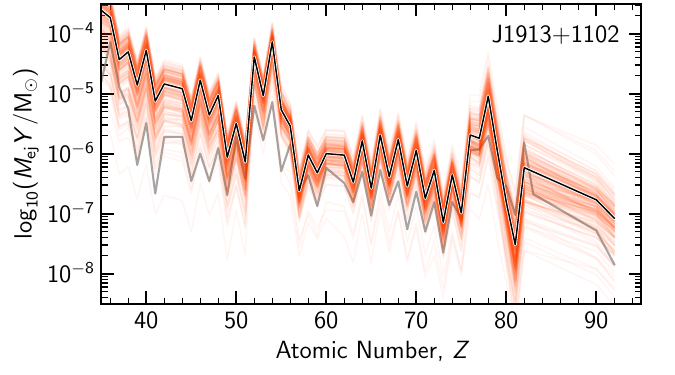}\\
	\includegraphics[width=\columnwidth]{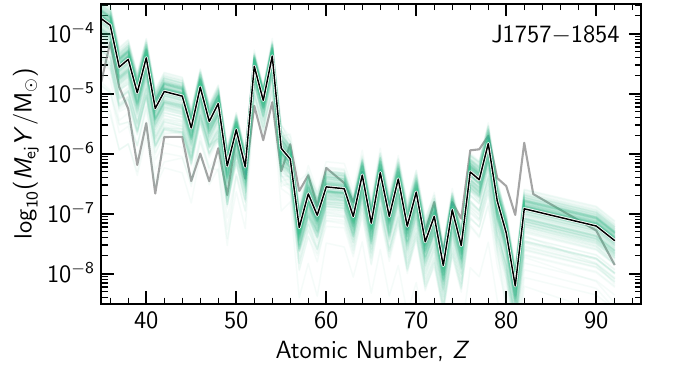}
	\includegraphics[width=\columnwidth]{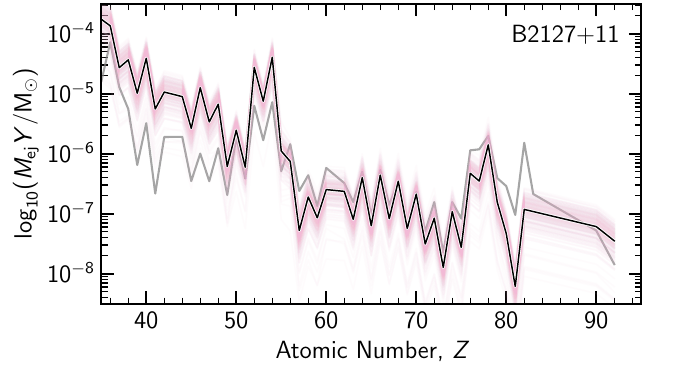}
    \caption{Predicted \rp\ yields of merger ejecta for DNS systems in the Galaxy compared to the \rp\ abundance pattern of the solar system, scaled by a constant (gray). Colored lines show individual random samples, while the black lines show the average between all draws. Note that {B2127+11} is a system found in a globular cluster (see text for details).\label{fig:MW}}
 	\end{figure*}
	
 	\begin{figure*}[th]
 	\centering
	\includegraphics[width=\columnwidth]{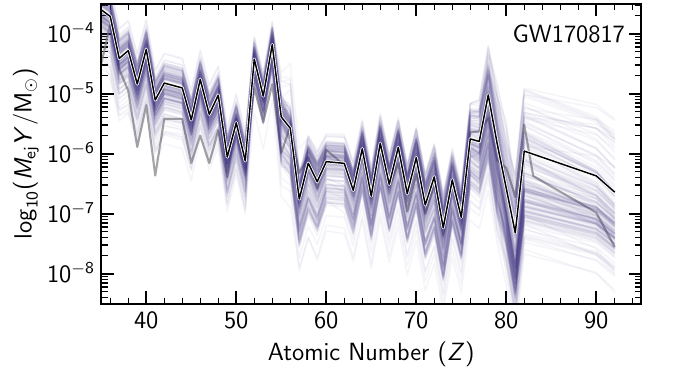}
	\includegraphics[width=\columnwidth]{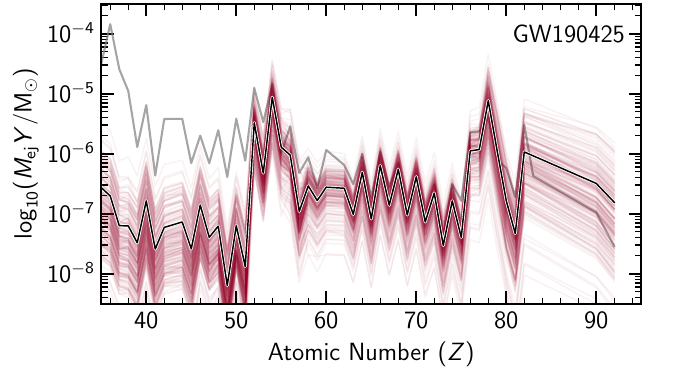}
    \caption{Predicted abundance patterns of merger ejecta for NSMs observed by LIGO/Virgo.\label{fig:GW_dd2}}
 	\end{figure*}

For each DNS, we put the 50 NS mass pairs through the separate equations that determine the dynamical and wind ejecta masses, the remnant lifetimes, and the dynamical ejecta centroid $Y_e$.
Next, 1000 samples are drawn randomly from normal distributions that describe the uncertainty associated with each of these four quantities.
Specifically, the uncertainty on the ejecta mass can be $>$50\% and have significant effect on the calculated yields.
Uncertainty on the abundances depends on the uncertainty in $\tau$ and $Y_e$ for the wind and dynamical ejecta, respectively.
For the disk lifetime (and therefore the disk ejecta mass), the median uncertainty is 100\%.
The centroid of the dynamical ejecta $Y_e$ is highly sensitive to the NS masses, significantly affecting the dynamical ejecta yields.
Therefore, for each DNS in Table~\ref{tab:results}, we propagate and quantify our systematic uncertainty by calculating 50,000 sets of yields.

\section{Results}
\label{sec:results}

For each DNS system modeled with our fiducial model, Table~\ref{tab:results} shows the median wind and dynamical ejecta masses, their respective lanthanide mass fractions, and the 16\textsuperscript{th} and 84\textsuperscript{th} percentile for each of these quantities, represented as upper and lower uncertainties.
Overall, these DNSs eject roughly equal amounts of matter owing to their similar masses.
Notably, the wind ejecta mass dominates over the dynamical ejecta mass by about a factor of 50.
In addition, note that the lanthanide mass fraction of the wind ejecta is 1.9$\times$10$^{-3}$ for almost all mass pairs.
This phenomenon can be attributed to the effect of the relatively stiff EOS on the remnant lifetime.
The remnant lifetime prescription adopted in our fiducial model \citep{lucca2020} is highly sensitive to the maximum non-rotating NS mass, $M_{\rm TOV}$.
The somewhat high $M_{\rm TOV}$ of the DD2 EOS ($\approx$2.4\,M$_{\odot}$) leads to lifetimes consistently above 530~ms, causing the NSM to eject its entire disk and result in lower neutron richness for the \rp\ in the wind ejecta (see, e.g., \citealt{Lippuner2017} and \citealt{holmbeck2021}).
Note, our fiducial model also therefore predicts that GW170817 was long-lived ($\tau>530$~ms), in agreement with more rigorous studies that find lifetimes for the GW170817 remnant $\mathcal O(1~{\rm s})$ \citep[e.g.,][]{Gill2019}. 
GW190425 is altogether different.
The significantly larger NS masses and higher mass asymmetry cause the NSM to eject very little material through the wind (three orders of magnitude lower than that of GW170817), though we note the wind has a large lanthanide mass fraction in our model.

In addition to the overall yields of the individual ejecta components, our method allows us to calculate projected detailed elemental abundance patterns.
Figure~\ref{fig:MW} shows the expected yields for each merging Galactic DNS system, as well as the DNS found in the globular cluster M15, {B2127+11}.
Each colored line shows one of the random samples drawn within the model uncertainties.
For plotting simplicity, we only show 200 randomly selected samples from the full set of 50,000 samples.
The black lines display the element-by-element averages calculated from the entire set of random draws.
The abundances are scaled by the total ejecta mass (see Section~\ref{sec:fiducial}), but note that they are not yet scaled by the weight.
Interestingly, the yields are directly comparable, and it can be directly seen that the systems ejecta roughly equal amounts of \rp\ material.
However, this total yield is difficult to directly compare to observational abundances due to the unknown mass in which this yield may be diluted.
Therefore, we also plot the Solar system \rp\ pattern (gray line) for reference, which we vertically shift by the same, arbitrary value for each system.

The low total masses and high symmetry values of the Galactic DNSs typically lead to strong production of the first \rp\ peak (e.g., $_{38}$Sr) relative to the rest of the heavy elements.
J1913$+$1102, the only DNS system in this sample with significantly unequal neutron star masses, produces the highest amount of actinides and the highest total ejecta mass overall.
The predicted abundances for the globular cluster system, {B2127+11}, behave similarly to the majority of DNSs in the Milky Way, though it has one of the lowest yields of actinide abundances.

Predicted abundance patterns for the DNSs detected by gravitational waves from their merger are shown in Figure~\ref{fig:GW_dd2}.
Recall that for these systems, the neutron star masses are drawn as co-dependent pairs from the (low-spin) posterior mass distributions, as described in Section~\ref{sec:method}.
This spread on the masses is an additional source of uncertainty on the abundance patterns and predicted ejecta masses, leading to a wider spread in the abundance patterns that are produced.
Interestingly, of all the systems, GW170817 shows the highest actinide abundance.
In addition, its total ejecta mass is among the largest of the DNSs in this sample, which---like J1913+1102---can be attributed to its slightly higher mass and asymmetry than DNSs in the Galaxy.
In contrast to all other systems we consider here, we find that the abundance pattern for GW190425 is suppressed in the first \rp-peak elements.
Additionally, the high NS masses and mass ratio of GW190425 is predicted to eject very little mass, though with very high lanthanide mass fraction, possibly due in part to its low first \rp-peak mass.
Compounded with its relatively large distance and poor sky localization, the low luminosity produced from this small amount of ejecta mass could in part explain why a kilonova counterpart could not be found \citep{antier2020, coughlin2019, hosseinzadeh2019, lundquist2019}.

\begin{figure}[t]
 	\centering
	\includegraphics[width=\columnwidth]{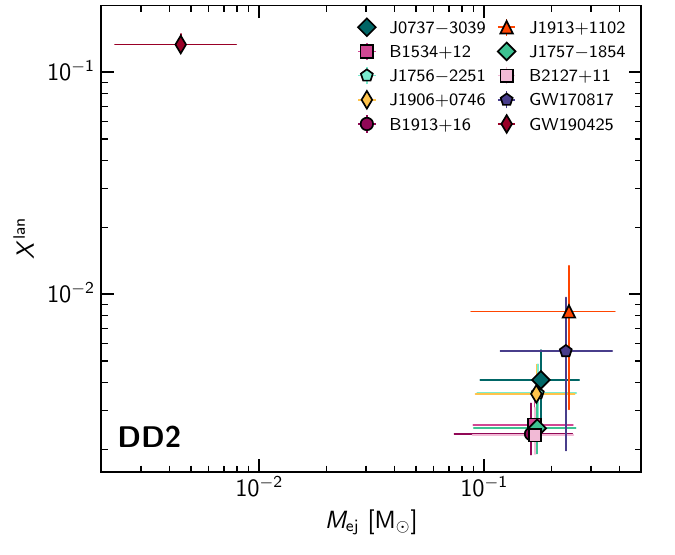}
	\caption{Total ejecta mass and lanthanide mass fractions for each DNS in Table~\ref{tab:results}.\label{fig:xlan}}
 	\end{figure}

The combined total ejecta masses and lanthanide mass fractions for the DNSs are displayed in Figure~\ref{fig:xlan}. For most of the DNSs, the total ejecta masses are consistently large ($\gtrsim$0.1~\Msun), while the lanthanide mass fractions are $\lesssim$10$^{-2}$. These DNSs are consistent with GW170817, in which our model predicts a total lanthanide mass fraction of $\simeq10^{-2.25}$, in agreement with several model studies of the AT2017gfo kilonova \citep{Arcavi2017,Chornock2017,Kilpatrick2017,McCully2017,Nicholl2017,Tanaka2017,Tanvir2017,Troja2017}. On the other hand, Figure~\ref{fig:xlan} highlights how our fiducial model predicts vastly different ejecta masses ($\lesssim10^{-2}$~\Msun) and lanthanide fractions ($\gtrsim$0.1) for {GW190425}.

\subsection{Fiducial model: total yields}
\label{sec:results_fiducial}

    \begin{figure}[t]
 	\centering
	\includegraphics[width=\columnwidth]{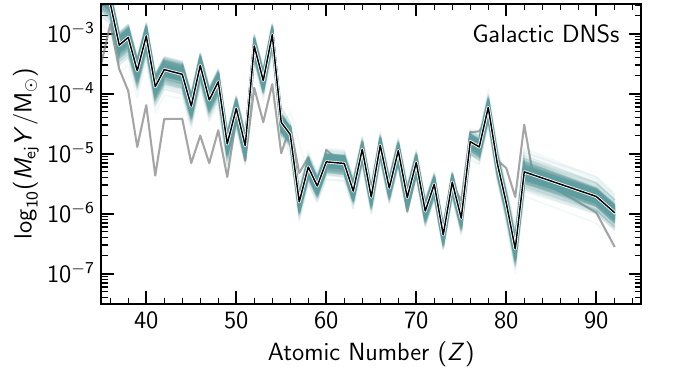}
    \caption{Predicted combined abundance patterns of merger ejecta for DNSs in the Galaxy.\label{fig:MW_combo}}
 	\end{figure}

With the individual yields for each Milky Way DNS calculated, we can also test the combined yields from these systems.
For this combination, we take the masses from the seven Galactic DNS systems and weigh their contribution by how representative each DNS is of similar-mass NSMs in population synthesis models.
The weights are listed in Table~\ref{tab:results}, and details of their calculation can be found in \citet{andrews2020}.
We multiply each system's total yield by their respective weight in Table~\ref{tab:results} and divide by the sum of the total weights.
As described in \citet{andrews2020}, the weight is essentially the expected merger rate (in Myr$^{-1}$) for each type of system.

Figure~\ref{fig:MW_combo} shows this combined, weighted abundance patterns of the seven Galactic DNSs.
Although low-mass, very symmetric DNS systems are the most common in the Milky Way sample, higher-mass systems like {B1913+16} are expected to merge more frequently and thus contribute more to the total abundance.
As in Figures \ref{fig:MW} and \ref{fig:GW_dd2} we compare the resulting abundance pattern to that of the Solar system in gray.
We again emphasize that the Solar system \rp\ pattern is representative of the relative pattern of each of element; the absolute vertical scaling is arbitrary, as there is no accurate way to convert the Solar system abundance pattern to an individual merger ejecta pattern.
Combined, the total merger yields from Milky Way DNS systems look similar to the Solar system \rp\ pattern, though with a higher enhancement to the lighter of the \rp\ elements.


\subsection{Macroscopic model uncertainties}
\label{sec:results_macro}

    \begin{figure*}[thb]
 	\centering
	\includegraphics[width=\columnwidth]{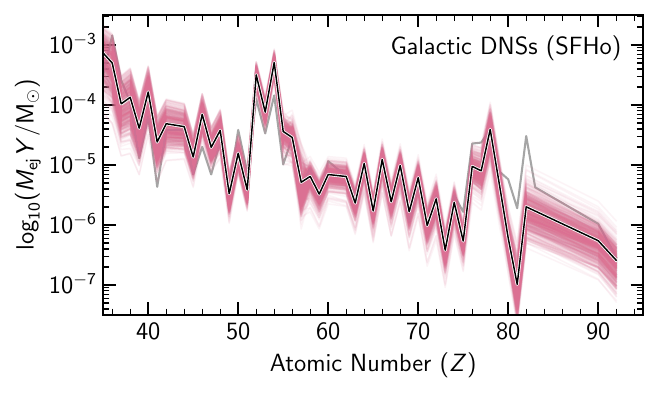}
	\includegraphics[width=\columnwidth]{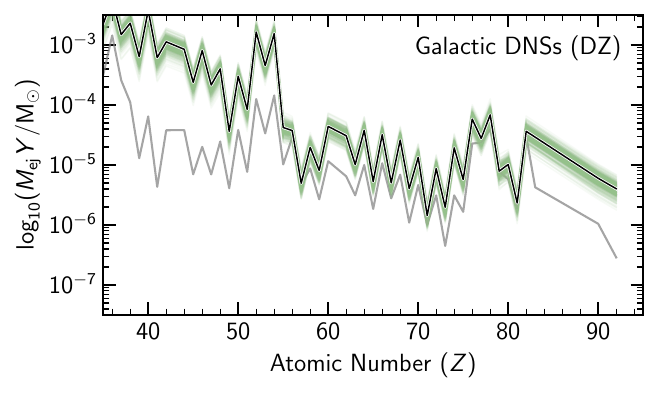}
    \caption{Effect on the combined abundance pattern of merger ejecta for DNSs in the Galaxy from a different EOS (left) and a different nuclear mass model (right).\label{fig:nuclear}}
 	\end{figure*}

We have chosen state-of-the-art nuclear reaction and decay rates, calculated self-consistently within the FRDM2012 nuclear model as well as some of the most up-to-date fits to hydrodynamic models of the disk and dynamical ejecta.
In addition, the composition of the disk ejecta was found by calculating the nucleosynthesis from tens of thousands of individual tracer particles from hydrodynamic simulations that vary the lifetime of the merger remnant.
Nevertheless, there are several additional areas of systematic uncertainty in this analysis.
For example, the dynamics of the accretion disk wind remain a topic of intense study with few to no constraints. 
Furthermore, other analytic descriptions of the dynamical and post-merger ejecta masses exist \citep[e.g.,][]{Dietrich2017,Radice2018,coughlin2019,dietrich2020,krueger2020,nedora2021}.
We test 25 unique combinations of dynamical and disk ejecta masses from the literature and found that there was very little sensitivity on the total combined abundance pattern.
In all model ejecta variations we tested, the long lifetime of the merger remnant caused the wind component to consistently dominate over the dynamical ejecta.
The largest variations we observed were that the dynamical ejecta mass described by \citet{Radice2018}, which leads to a slightly less-abundant third \rp\ peak, and the description by \citet{Dietrich2017}, which produces a higher third peak.
However, we find the uncertainty from the disk and dynamical ejecta descriptions to be minor overall, especially compared with the microscopic model assumptions which we discuss in the subsequent section.

\subsection{Microscopic model uncertainties}
\label{sec:results_micro}

\begin{figure}[h!]
 	\centering
	\includegraphics[width=\columnwidth]{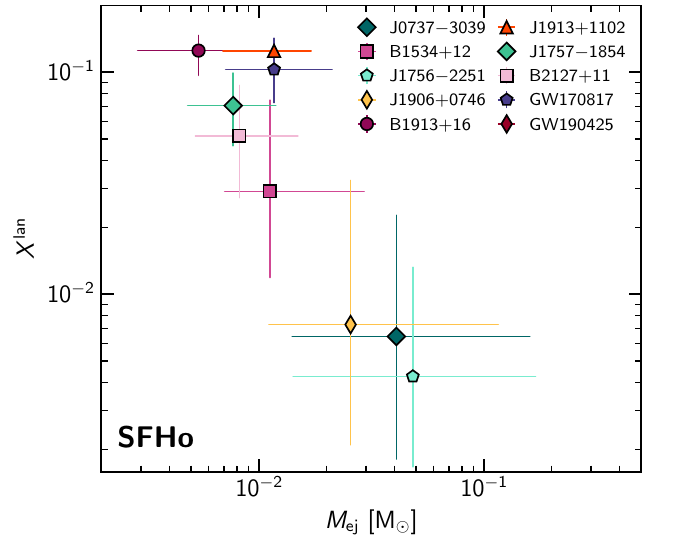}
	\caption{Total ejecta mass and lanthanide mass fractions using the SFHo EOS. Note that GW190425 is off the scale of this plot near $M_{\rm ej}\sim 10^{-4}$ and $X_{\rm lan}\sim 10^{-1}$.\label{fig:xlan_SFHo}}
 	\end{figure}

While the nucleosynthetic yield is relatively insensitive to the exact choice of disk and dynamical ejecta models for our fiducial model, these variations become important when changing the EOS and nuclear mass model.
To demonstrate these effects, we choose a softer EOS, SFHo \citep{Steiner2013}, and show the combined Milky Way abundances from the seven Galactic DNSs in the left panel of Figure~\ref{fig:nuclear} and report the results in Table~\ref{tab:results_sfho}.
The lower $M_{\rm TOV}$ and smaller NS radii given with SFHo leads to much smaller wind ejecta masses (ranging from a factor of five to 500) and dynamical ejecta masses consistently about two times larger then the results in Table~\ref{tab:results}.
The lanthanide mass fractions also vary for the components.
Generally with the SFHo model, the wind ejecta can reach higher neutron richness (allowed by lower remnant lifetimes), with a maximum lanthanide mass fraction of $10^{-1.32}$ ({B1913+16}).
On the other hand, the average $Y_e$ for the dynamical ejecta is predicted to be somewhat higher with SFHo, causing the dynamical ejecta lanthanide mass fraction to be overall comparable to, but slightly lower than, our fiducial EOS choice (DD2).
The effect of the softer EOS is that first and second \rp-peak production is somewhat decreased, leading to a more consistent overall fit with the Solar \rp\ abundances.
We summarize the combined ejecta masses and lanthanide mass fractions with the SFHo model in Figure~\ref{fig:xlan_SFHo}.
Note that there is more variation in the ejecta masses and lanthanide production when using SFHo, unlike in Figure~\ref{fig:xlan} in which the calculations tended to group around the same values for nine of the ten DNS systems.

Next, we run all of our nucleosynthesis calculations with the Duflo-Zuker nuclear mass model \citep[DZ;][]{Duflo1995} instead of our baseline FRDM2012 model.
For this test, we revert back to the DD2 EOS.
Note that changing the nuclear mass model does not affect the total NSM ejecta mass.
This change required recalculating thousands of thermodynamic trajectories, which we are able to accomplish using the PRISM code.
The right panel of Figure~\ref{fig:nuclear}, shows the independent effect of changing the nuclear mass model on the combined abundance pattern.
The net effect of employing a different nuclear mass model for the composition of the ejecta is that the lanthanides are produced in higher amounts and the actinides are slightly reduced.
In addition, the \rp\ peaks have lower abundances relative to the baseline case in which FRDM2012 nuclear masses are used.


\section{Discussion and Conclusions}
\label{sec:conclusions}

In this work, we have presented our model predictions for the \rp\ abundances from NSMs inside and outside of the Galaxy.
We find that the total contribution by Galactic NSM systems produce overall lower actinide abundances compared to gravitational-wave detected NSMs (i.e., compare Figures~\ref{fig:MW_combo} and \ref{fig:GW_dd2}).
In addition, our predictions for \rp\ nucleosynthesis from Milky Way DNSs fail to match the Solar actinide abundances, which may imply that DNSs are not sufficiently frequent or high-yield to account for what we see in the Galaxy on average. We have demonstrated that this conclusion is robust to reasonable changes to choices for the macroscopic model, i.e., changes to the disk wind and dynamical eject models. On the other hand, the microscopic model uncertainties produce significant changes to the resulting elemental yields. When using a softer EOS, SFHo, we find an elemental abundance pattern generated by NSMs more closely matching the Solar pattern, but a lanthanide fraction for GW170817 too high compared with observations.
Similarly, employing a different nuclear mass model (DZ) for the nucleosynthesis calculations produced overall higher yields of the first and second \rp\ peaks.

\begin{deluxetable*}{l c c c c c c c}
\tablecaption{Observed DNS systems and their ejecta properties as in Table~\ref{tab:results}, but for the SFHo EOS.\label{tab:results_sfho}}
\tablehead{
\colhead{Name} & \colhead{$M_1$} & \colhead{$M_2$} & \colhead{Weight} & \colhead{$m_{\rm wind}$} & \colhead{$m_{\rm dyn}$} & \colhead{$X^{\rm lan}_{\rm wind}$} & \colhead{$X^{\rm lan}_{\rm dyn}$}
\\
\colhead{} & \colhead{(M$_\odot$)} & \colhead{(M$_\odot$)} & \colhead{} & \colhead{(10$^{-3}$\,M$_\odot$)} & \colhead{(10$^{-3}$\,M$_\odot$)} & \colhead{($\times$10$^{-3}$)} & \colhead{($\times$10$^{-3}$)}
}
\startdata
\multicolumn{8}{c}{Field Systems}\\
\hline
J0737$-$3039	& 1.34 	& 1.25 & 4.8 & $35.6_{-26.7}^{+119.1}$ & $5.3_{-2.5}^{+2.6}$ & $1.9_{-0.2}^{+9.2}$ & $37.0_{-23.4}^{+33.7}$\\
B1534+12	    & 1.35	& 1.33 & 0.6 & $6.0_{-3.3}^{+18.0}$ & $5.2_{-2.5}^{+2.6}$ & $7.5_{-5.7}^{+47.7}$ & $53.9_{-31.5}^{+30.9}$\\
J1756$-$2251	& 1.31	& 1.26 & 0.6 & $43.5_{-34.1}^{+121.3}$ & $4.9_{-2.3}^{+2.4}$ & $1.9_{-0.2}^{+6.3}$ & $25.2_{-16.9}^{+34.1}$ \\
J1906+0746	    & 1.32	& 1.29 & 2.2 & $20.6_{-14.3}^{+90.4}$ & $5.0_{-2.4}^{+2.4}$ & $2.0_{-0.2}^{+19.1}$ & $29.1_{-19.2}^{+34.5}$ \\
B1913+16	    & 1.44	& 1.39 & 6.5 & $0.3_{-0.1}^{+0.2}$ & $5.1_{-2.5}^{+2.5}$ & $47.6_{-21.2}^{+21.8}$ & $129.6_{-29.8}^{+22.0}$ \\
J1913+1102	    & 1.62	& 1.27 & 0.5 & $1.7_{-0.8}^{+1.4}$ & $10.0_{-4.7}^{+5.2}$ & $45.8_{-8.2}^{+21.1}$ & $138.1_{-9.3}^{+11.8}$ \\
J1757$-$1854	& 1.40	& 1.34 & 7.8 & $2.3_{-1.2}^{+3.3}$ & $5.4_{-2.6}^{+2.6}$ & $30.2_{-27.8}^{+35.7}$ & $87.9_{-29.8}^{+26.9}$ \\
\hline
\multicolumn{8}{c}{Globular Cluster System}\\
\hline
B2127+11	& 1.36	& 1.35 & --- & $3.0_{-1.6}^{+6.2}$ & $5.2_{-2.5}^{+2.6}$ & $18.1_{-16.2}^{+44.3}$ & $71.0_{-34.7}^{+27.8}$ \\
\hline
\multicolumn{8}{c}{Gravitational Wave Detections}\\
\hline
GW170817	& $1.56\pm0.11$ & $1.22\pm 0.09$ & --- & $3.6_{-2.2}^{+5.9}$ & $8.1_{-4.0}^{+7.4}$ & $40.5_{-36.8}^{+27.2}$ & $130.4_{-33.4}^{+19.6}$ \\
GW190425	& $2.10\pm 0.31$ & $1.43\pm0.17$ & --- & $0.33_{-0.27}^{+1.38}$ & $0.03_{-0.02}^{+0.04}$ & $37.0_{-0.0}^{+0.0}$ & $104.2_{-5.9}^{+12.5}$ \\
\enddata
\end{deluxetable*}

While we have not exhaustively examined the entire range of possible model variations, none of our models can simultaneously fit the entire Solar \rp\ abundance pattern.
With either a different EOS describing the ejecta or a different nuclear mass model affecting the composition, the actinides are consistently underproduced relative to the Solar system lanthanides. 
This inconsistency suggests that a contribution from events that produce higher actinide abundances may be needed: either more asymmetric NSMs, a contribution from neutron-star black-hole mergers, or other sites that were active in the Galaxy's early history.

In addition to the multiple model uncertainties we adopt, our approach intrinsically makes several assumptions in extrapolating to the entire \rp\ production history of the Milky-Way.
First, when we combine our individual NMS yields into one ``Galactic DNS" abundance pattern, we inherently assume that the ejecta has mixed homogeneously into the interstellar medium (ISM) and in proportion to both the total ejecta mass and the weight ascribed to each DNS.
Fully capturing the subtleties of NSM rates and yields and how those quantities interact with star-formation demands a full galactic chemical evolution analysis, which is beyond the scope of this work.
Our combined NSM yields in Figure \ref{fig:MW_combo} can instead be thought of as a representation of the average Milky Way \rp\ contribution by NSMs rather than a robust prediction.

Second, the Galactic DNSs may not be a representative sample of a population of merging neutron-star binaries.
The pulsars in our sample of Galactic DNSs in Table~\ref{tab:results} all have relatively young spin-down ages \citep{tauris2017}, suggesting they were made from metal-rich stars.
Massive metal-poor binaries, like those seeding the Milky Way with \rp\ elements early in its formation, may produce different mass neutron stars and therefore a different elemental abundance pattern upon merger.
Additionally, the detection of GW190425 confirms that neutron star binaries more massive than those currently observed in the Milky Way sample can exist. 
Although its formation is yet to be uncovered, two possibilities have been presented in the literature to explain the mass discrepancy between this system and other Galactic DNSs; either a formation channel produces DNSs at very short orbital periods that they quickly merge, making them unlikely to be observed \citep[e.g.,][]{romero-shaw2020}, or some process (such as mass fallback, burying their magnetic fields) exists that keeps the NSs in some systems from becoming pulsars, or at least forms pulsars with a short lifetime \citep{sarafzadeh2020, vigna_gomez2021}. 
Either possibility suggests that the observed sample listed in Table~\ref{tab:results} forms an incomplete, or ``hidden," representation of the population of merging DNSs.
Nevertheless even if a hidden population exists, Figure~\ref{fig:xlan} shows that massive NSMs such as GW190425 produces at least an order of magnitude less ejecta mass compared with the other systems. 
Assuming that the rate of NSMs from any putative hidden population does not dwarf the merger rate of those systems observed in the Milky Way, we can be optimistic that any subpopulation of GW190425-like systems missing from the current Milky Way census would provide at most a perturbation to the integrated Milky Way abundance pattern. 

Finally, any comparison between the model data and the Solar values inherently assumes that NSMs are the main source of elements from the first \rp-peak to the actinides.
We have neglected additional sources from, e.g., exotic SNe that may produce a robust \rp\ pattern.
These sites are thought to be roughly ten times more frequent than NSMs, but may eject ten times less the material that an NSM would \citep[e.g.,][]{Tsujimoto2014,cote2017}.
Similarly, we have also neglected the contribution by neutron-star black-hole mergers.
These types of mergers may produce a ``dynamical-only" abundance pattern that can serve to dilute the strong first-peak production of the NSM-only abundances in this work \citep[see, e.g., recent work by][]{Wanajo2022}.
Such compact object mergers may also be ten times less frequent than NSMs, but could ejecta ten times as much \rp\ material.

Each of the sources of uncertainty in this model, from the microphysics of nuclear reactions to the hydrodynamics of NSM ejecta, has their own dedicated studies detailing how each piece of physics affects the resulting \rp\ yields.
This study is presented as a first step towards visualizing how a population of DNSs propagates into a combined, Galactic \rp\ abundance pattern.

\acknowledgements
This work was initiated at the Aspen Center for Physics, which is supported by National Science Foundation grant PHY-1607611. We would especially like to thank the organizers of the 2021 summer workshop ``Galactic Archaeology with Fundamental Stellar Parameters: Synthesizing the Power of Accurate Ages with Distances, Chemistry, and Kinematics," during which this work was initiated.
Support for this work was provided by NASA through the NASA Hubble Fellowship grant HST-HF2-51481.001 awarded by the Space Telescope Science Institute, which is operated by the Association of Universities for Research in Astronomy, Inc., for NASA, under contract NAS5-26555 (EMH).
JJA acknowledges support from CIERA and Northwestern University through a Postdoctoral Fellowship.
EMH and JJA additionally acknowledge support by a grant from the Simons Foundation.


\bibliographystyle{aasjournal}
\bibliography{bibliography.bib}

\end{document}